# TEST OF ACCELERATING STRUCTURE FOR VEPP-5 PREINJECTOR

V.E. Akimov, A.V. Aleksandrov, A.V. Antoshin, M.S. Avilov, P.A. Bak, O.Yu. Bazhenov,
Yu.M. Boimelshtein, D.Yu. Bolkhovityanov, A.G. Chupyra, N.S. Dikansky, A.R Frolov,
R.Kh. Galimov, R.G. Gromov, K.V. Gubin, S.M. Gurov, Yue.A. Gusev, I.V. Kazarezov,
V.D. Khambikov, S.N. Klyuschev, N.A. Kiseleva, V.I. Kokoulin, M.B. Korabelnikov,
A.A. Korepanov, V.I. Kopylov, A.N. Kosarev, N.Kh. Kot, N.N. Lebedev, P.V. Logatchev,
A.N. Lukin, P.V. Martyshkin, L.A. Mironenko, A.A. Nikiforov, A.V. Novokhatski, V.M. Pavlov,
I.L. Pivovarov, O.V. Pirogov, V.V. Podlevskikh, S.L. Samoilov, V.S. Severilo, V.D. Shemelin,
S.V. Shiyankov, B.A. Skarbo, A.N. Skrinsky, B.M. Smirnov, A.N. Sudarkin, D.P. Sukhanov,
A.S. Tsyganov, Yu.V. Yudin. N.I. Zinevich, Budker INP, 630090, Novosibirsk, Russia

*Abstract*

Preinjector VEPP-5 electron linac consists of two linear accelerators on energy 300 MeV and 510 MeV and includes 14 accelerating structures [1]. First accelerating structures both linacs have increased average rate of acceleration 25-30 MeV/m, and other regular sections up to 17-20 MeV/m. This paper presents the results of basic test of 3-meter long accelerating structure. Average rate of acceleration of an electron beam 35 MeV/m was achieved. The electron beam energy up to 105 MeV was obtained.

## 1 INTRODUCTION

Tests were carried out on the basis of initial part of preinjector (first accelerating assembly) that includes the electron gun, subharmonic buncher, S-band buncher, three accelerating sections, RF assembly on the basis of 5045 klystron, system of power compression (SLED-type), focusing system and a system of beam diagnostics (see Fig. 1). To get the maximum accelerating ratio, all RF power after the compression is directed into the first accelerating section. The rest two sections and the subharmonic buncher were out of operation in present experiments and were used as the channel of beam transportation.

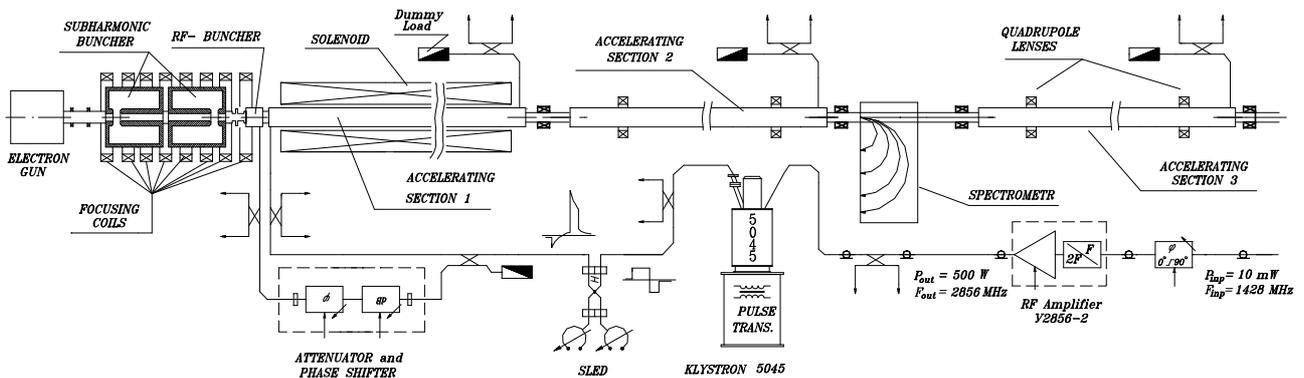

Figure 1: The scheme of RF test accelerating section.

## 2 ACCELERATING SECTION

The accelerating section (AS) of VEPP-5 preinjector is managed as the cut of a round disk-loaded waveguide with the constant impedance (the constant cell geometry along the AS). Two wave type transformers (WTT) are at the input and the output of the section. WTT transforms the basic mode of waveguide channel $H_{10}$ into the accelerating mode $E_{01}$ of a round disk-loaded waveguide. To reduce the overvoltage ratio, the edge of the iris is performed as an ellipse with the half-axis ratio 1:2.

The section was assembled of two identical part 1.5 m length each. Brazing of each part was carried out in the vacuum furnace. Both brazed half-sections were welded together via the connecting iris, as shown in Fig. 2. Both sides of connecting iris were covered by the Au film. The iris was tightening between the special connecting cells up to formation of the thermal diffusion seam. Then, in order to provide the mechanical strength, the welding of steel rings was done.

The AS parameters are presented in Table 1.

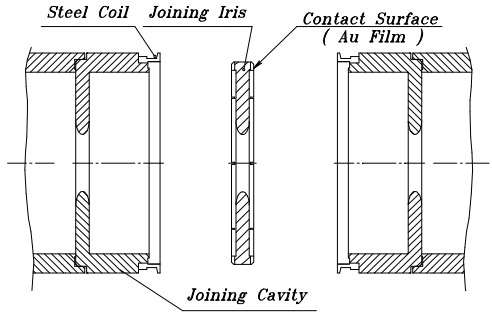

Figure 2: Connection of two AS parts.

Table1: Parameters of the accelerating section

| | |
|---|---|
| Operational frequency | 2855.5 MHz |
| Internal cell diameter $2b$ | 83.75 mm |
| Iris diameter $2a$ | 25.9 mm |
| Iris thickness $t$ | 6 mm |
| Period $D$ | 34.99 mm |
| Operational mode of oscillation $\theta$ | $2\pi/3$ |
| Relative phase velocity $\beta_p$ | 1 |
| Relative group velocity $\beta_g$ | 0.021 |
| Section length $L$ | 2.93 m |
| Total number of cells (incl. 2 WTT) | 85 |
| Unloaded quality factor $Q_0$ | 13200 |
| Shunt impedance $R_{sh}$ | 51 MOhm/m |
| Time constant $\tau_{0a}=2Q_0/\omega_0$ | 1.471 μs |
| Attenuation (by field) $\alpha=1/(\tau_{0a}v_{gr})$ | 0.108 m$^{-1}$ |
| Filling time $T_f=L/v_{gr}$ | 0.465 μs |

Before the experiments, the AS and all waveguide channel were heated for 3 days at the temperature $t \approx$ 230°C. After the heating and RF training, the vacuum of $\sim 3 \cdot 10^{-8}$ Torr was achieved.

## 3 RF POWER SUPPLY

The continuous RF signal ($f$ = 1428 MHz, $P \approx$ 10 mW) from the master oscillator is moved to the forming amplifier assembly U2856-2 via 90° phase shifter. The forming amplifier operates at the doubled frequency 2856 MHz and forms the RF pulse of ~3.5 μs width and output pulsed power 200÷400 W for 5045 klystron excitation. Phase shifter provides the phase inversion at the operational frequency at the given time to ensure the pulse compression system operation. RF pulse of up to 60 MW power is moved from the klystron along the waveguide channel (72x 34 mm$^2$) via the SLED-type pulse compression system, to the AS input. The unused power in the AS is dissipated in the matched load that is located at the output of the section. The fraction of RF power after SLED system is directed to the RF buncher via the direct coupler with the attenuation 23 dB. RF pulse phase and amplitude can be varied by means of the attenuator and phase shifter.

The control of work temperature of AS, load, RF buncher and SLED cavities is provided by the system of thermostabilization. The stabilization of temperature is ensured by change of temperature of cooling distilled water at the input of cooling systems. Operational temperature of AS is 30°÷35°C, accuracy of temperature stabilization is ±0.1°C.

## 4 RESULTS OF MEASUREMENTS WITH THE SHORT BUNCH

During the experiments the main gun parameters were:
- Electron gun voltage   170 kV,
- Gun pulsed current     3 A,
- Current pulse width    2 ns,
- Repetition rate        5 ÷50 Hz.

For this pulse width the beam radiation field is low and it doesn't influence essentially on the acceleration. The accelerating field in the AS is defined by the klystron input power only. At that operational conditions it is possible to reach the maximum accelerating gradient and, hence, the maximum energy of the accelerated beam.

The operational parameters of power compression system cavities are calculated, using the method of regressive analysis and the RF pulse shape after the power compression system:
- Unloaded quality factor   $Q_0 = 7.58 \cdot 10^4$,
- Coupling ratio            $\beta = 8.2$,
- Time constant             $\tau_0 = 8.45$ μs,

and also typical times of input signal transient processes:
- $\tau_1 = 0.021$ μs - typical time of growth of incident
- wave amplitude leading pulse edge,
- $\tau_2 = 0.044$ μs - typical time of phase inversion.

As an example, Fig. 3 shows the pulse shape after the power compression system at the pulsed input power amplitude 50 MW.

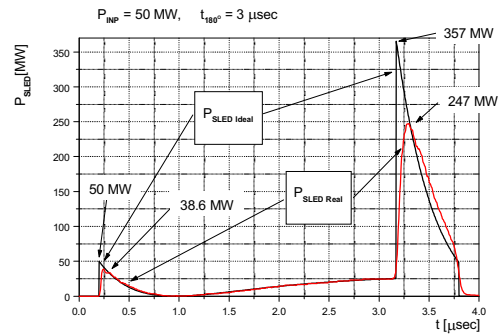

Figure 3: Pulse shape after pulse comression system.

One can calculate the distribution of accelerating electric field amplitude along the AS at different times by the shape of RF pulse at the AS input and AS parameters. For constant impedance structure:

$$E(z,t) = E_0\left(t - \frac{z}{v_{gr}}\right)e^{-\alpha z},$$

where $E_0(t) = E(0,t)$ is electric field amplitude at the AS input at $z = 0$. Using the accelerating field distribution one can calculate the energy gain of the accelerated particles that enter the AS at various times.

Gun pulsed current is measured by the resistance monitor located at the gun output. Measurements of beam energy are made by means of 180° magnetic spectrometer. After spectrometer the beam was directed to the luminescent screen. One can define the energy of accelerated electrons by the value of magnetic field in the spectrometer and radius of turn. Also the beam energy spread can be evaluated by the size and relative brightness of the spot on the luminescent screen. The charge of accelerated beam is measured by the Faraday cup, located after the c.

At present time at the operational conditions of short beam at the first AS output the accelerated beam with $2 \cdot 10^{10}$ particles in the bunch is obtained. Maximum average accelerated gradient 35 MeV/m is reached. The maximum amplitude of accelerating field in first cells is 50 MV/m.

The AS operated at 5 Hz repetition rate (output energy of electrons 106 MeV) with a single breakdown during 40 minutes of operation, and without any breakdown at 50 Hz repetition rate (output energy 75 MeV). At the operation with RF buncher switched on the energy spread in the beam was ±0.5 %.

## 5 RESULTS OF MEASUREMENTS WITH THE LONG BUNCH

The goal of these experiments was the AS test at the conditions operating of the accelerator for project IREN [2]. Test conditions were:
- Repetition rate          1÷5 Hz,
- Electron gun voltage     170 kV,
- Gun pulsed current       2.6 A,
- Current pulse width      310 ns,
- Output klystron power    ≈ 50 MW,
- RF pulse duration        3.6 μs,
- Time before phase inversion 2.38 μs.

In that experiment RF buncher wasn't fed by the power and unbunched beam was steered to the AS input, that is why the accelerated beam had large energy spread. The photograph of luminescent screen after 180° magnetic spectrometer is shown in Fig. 4.

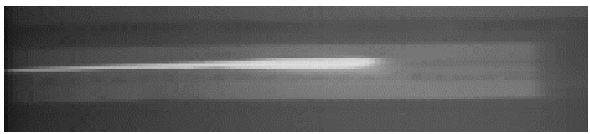

Figure 4: Photograph of accelerated beam from luminescent screen.

At these conditions the AS operates with strong current load that leads to essential change of the AS accelerating field and accelerated beam characteristics. These conditions are typical for the beam with high-energy capacity. Besides, the level of RF power transferred to the load after the AS (see Fig. 5) was measured with and without beam load. Also current pulse profiles from the gun, obtained from the wall current monitor, are shown in that figure.

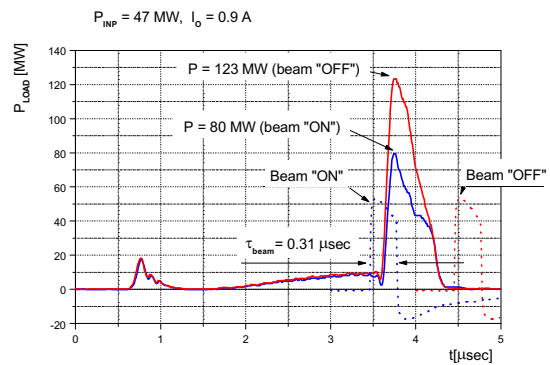

Figure 5: The shape of RF power transferred to the load with and without beam load. The signals of current pulse from the gun wall current monitor.

The main results are:
- Maximum electron energy         92 MeV,
- Minimum electron energy         47 MeV,
- The total charge of the beam    $2.8 \cdot 10^{-7}$ C,
- (total number of particles      $1.75 \cdot 10^{12}$),
- The beam energy content         26.1 Joile.

The details of the experiments with long bunch are in joint report presented on this conference [3].

## REFERENCES

[1] A.V. Alexandrov et. al., "Electron-positron preinjector of VEPP-5 complex". Proceedings of the XVIII International Linear Accelerator Conference., 26-30 August 1996, Geneva, Switzerland., CERN 96-07, 15 November 1996, Vol. 2, pp. 821-823.

[2] V. Antropov et. al., "IREN test facility at JINR". Proceedings of the XVIII International Linear Accelerator Conference, 26-30 August 1996, Geneva, Switzerland, Vol. 2, pp. 505-508.

[3] S.N. Dolya, W.I.Furman, V.V. Kobets, E.M. Lasiev, Yu.A. Metelkin, V.A. Shvets, A.P. Soumbaev, the VEPP-5 Team, "Linac LUE200. First testing results" (this conference).